%% file: main.tex
\documentclass[]{relaxed_system_lab}

\usepackage[utf8]{inputenc}
\usepackage[T1]{fontenc}
\usepackage{geometry}
\usepackage{amsmath,amssymb,amsfonts}
\usepackage{charter}
\usepackage{bm}

\usepackage{graphicx}
\usepackage{xcolor}
\usepackage{float}
\usepackage{wrapfig}
\usepackage{booktabs}
\usepackage{multirow}
\usepackage{array}
\usepackage{tabularx}
\usepackage{pgfplots}
\usepackage{pgfplotstable}
\usetikzlibrary{patterns}

\usepackage{minitoc}
\usepackage{appendix}
\usepackage{parskip}
\usepackage{soul}
\usepackage{textcomp}
\usepackage{tcolorbox}

\usepackage{xurl}
\usepackage{hyperref}
\usepackage{cleveref}   

\usepackage{caption}
\usepackage{subcaption}

\usepackage{algorithmic}

\usepackage{CJKutf8}
\usepackage{url}

\newcommand{\sys}{\textsc{Hexgen-Flow}\xspace}
\newcommand{\fred}[1]{{\color{black}{#1}}}

\definecolor{cvprblue}{rgb}{0.21,0.49,0.74}
\definecolor{fallbackgreen}{rgb}{130, 180, 102}
\definecolor{stopred}{rgb}{251, 225, 224}

\ifdefined\final
\usepackage[disable]{todonotes}
\else
\usepackage[textsize=tiny]{todonotes}
\fi

\input{macro}

\newtcolorbox{promptbox}[1][]{
  enhanced,
  breakable,
  colback=promptboxlightgray,
  colframe=promptboxblue!30,
  arc=8pt,
  boxrule=0.5pt,
  left=12pt,
  right=12pt,
  top=8pt,
  bottom=8pt,
  fonttitle=\bfseries,
  fontupper=\linespread{1.2}\selectfont,
  title=#1
}

\title{\sys: Optimizing LLM Inference Request Scheduling for Agentic Text-to-SQL}

\author{You Peng$^1$$^*$, Youhe Jiang$^1$$^*$, Wenqi Jiang$^2$, Chen Wang$^3$$^\dagger$, Binhang Yuan$^1$$^\dagger$}

\affiliation{$^1$HKUST, $^2$ETH Zurich, $^3$Tsinghua University}

\contribution{$^*$Equal contribution, $^\dagger$Corresponding author}

\abstract{
Recent advances in agentic large language models (LLMs) have substantially improved Text-to-SQL, enabling users without database expertise to query databases intuitively. However, deploying agentic LLM-based Text-to-SQL systems in production remains challenging due to multi-stage dependencies, strict latency requirements, and deployment complexity across heterogeneous GPUs in enterprise clusters. Existing LLM serving frameworks are designed mainly for independent inference tasks, leading to suboptimal performance and frequent service-level objective (SLO) violations for Text-to-SQL workloads.
In this paper, we introduce \sys, a framework for scheduling and executing agentic multi-stage LLM-based Text-to-SQL workflows on heterogeneous GPU clusters serving multi-tenant requests. \sys adopts a hierarchical scheduler that combines global workload-balanced task dispatching with an adaptive local priority queue, guided by a systematic analysis of agentic Text-to-SQL workflows. We also propose a lightweight simulation-based method to tune key scheduling hyperparameters, improving robustness and adaptability.
Evaluations on realistic Text-to-SQL benchmarks show that \sys significantly outperforms state-of-the-art LLM serving frameworks. Across all traces, \sys reduces P95 tail latency by $1.42{\sim}1.56\times$ and increases throughput by $1.49{\sim}1.81\times$, demonstrating consistent gains under diverse workloads.
}

\begin{document}

\maketitle

\section{Introduction}
\label{sec:intro}

Recent advances in large language models (LLMs) have substantially improved Text-to-SQL, i.e., translating natural language questions into executable SQL statements~\cite{gao2024text,li2024dawn,changdr,fu2023catsql}. This progress can democratize database access by enabling non-experts to query structured data without writing SQL. In enterprise settings, agentic Text-to-SQL can improve analytics productivity and accessibility.

\begin{figure}
    \centering
    \includegraphics{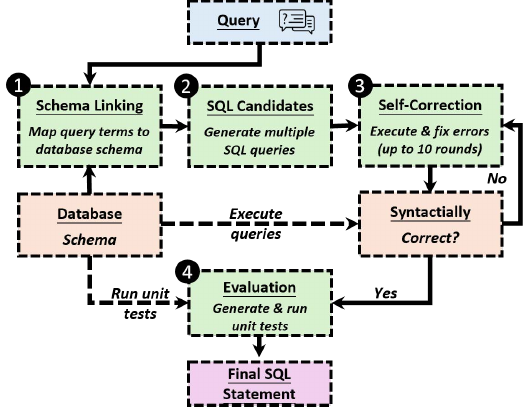}
    \caption{A visual illustration of the agentic multiple-stage Text-to-SQL workflow with inter-stage dependency.}
    \label{fig:t2qworkflow}
    \vspace{-1.5em}
\end{figure}

A typical agentic Text-to-SQL workflow (Figure~\ref{fig:t2qworkflow}) consists of multiple interdependent stages: (1) \textit{schema linking} to map query terms to the database schema; (2) \textit{SQL candidate generation} to propose multiple SQL statements; (3) \textit{self-correction} to execute and iteratively refine erroneous candidates; and (4) \textit{evaluation} to select a final statement using LLM generated unit tests. \fred{Compared to non-agentic Text-to-SQL methods with a small, fixed number of LLM calls, this agentic pipeline improves quality via verifier-guided refinement and candidate selection, but introduces a longer, dependency-constrained sequence of LLM calls. Unlike many general agentic workflows, it typically runs without human-in-the-loop, and \phantomsection\label{rev:sql_negligible} SQL execution is on the order of milliseconds---negligible relative to end-to-end latency dominated by multi-stage LLM inference.}

Deploying agentic LLM-based Text-to-SQL systems in production requires not only strong models~\cite{gpt4o,reid2024gemini,claude3,jiang2024mixtral,touvron2023llama} and inference-time scaling algorithms~\cite{snell2024scaling,brown2024large}, but also an efficient and scalable serving infrastructure. In practice, enterprise environments often rely on heterogeneous GPUs for LLM serving\footnote{In the current exciting era of generative AI, chip vendors typically release new generations of AI chips every 24 months. For example, Nvidia introduced the Turing architecture in 2018~\cite{Nvida_turing}, Ampere in 2020~\cite{Nvida_ampere}, Hopper in 2022~\cite{Nvida_hopper}, and Blackwell in Q4, 2024~\cite{Nvida_blackwell}. On the other hand, one particular version of an AI chip often serves for a much longer period.}. Therefore, the goal is to serve the multi-stage Text-to-SQL pipeline efficiently on heterogeneous clusters, achieving low end-to-end latency and high throughput.

However, efficient serving is challenging due to the workflow dependencies and the heterogeneity in both request demands and GPU serving capacities. Each end-to-end query triggers a sequence of dependent LLM calls that must be coordinated across model instances with different performance and load. The combination of workflow dependencies, resource heterogeneity, and diverse per-query service-level objectives (SLO) complicates scheduling and resource allocation. Concretely, we summarize three core challenges below:

\begin{itemize}
\item \textbf{\underline{C1.} LLM inference request dependencies}: Agentic Text-to-SQL workflows involve multiple interdependent stages with different urgency levels. Later-stage tasks cannot commence until the preceding stages are completed. Consequently, delays in early stages affect downstream stages, increasing the risk of end-to-end SLO violations.

\item \textbf{\underline{C2.} Heterogeneity in workflow and infrastructure}: The heterogeneity in Text-to-SQL workflow lies in two aspects: (\underline{i}) The LLM inference requests exhibit substantial variability in different stages, driven by differences in the length of the query prompt and the number of output tokens generated; (\underline{ii}) Enterprise production environments commonly leverage heterogeneous GPUs with different computational capabilities, so the latency/throughput of the same request can differ across model instances.

\item \textbf{\underline{C3.} Varying SLO constraints in serving}: Deployments must handle concurrent Text-to-SQL queries from different users with different SLOs. Ensuring that all LLM inference requests within a query collectively meet its end-to-end SLO requires fine-grained, per-query deadline awareness and adaptive prioritization across stages and model instances.
\end{itemize}

Due to these challenges, scheduling policies in popular LLM serving systems~\cite{kwon2023efficient,sheng2024fairness,patke2024QLM} are often inefficient for Text-to-SQL workloads. Existing frameworks largely target independent inference requests and lack explicit coordination across dependent stages or enforcement of end-to-end deadlines. While recent works on adaptive batching~\cite{agrawal2024taming}, priority-aware allocation~\cite{patke2024queue,wang2024revisiting}, and GPU load balancing~\cite{jain2025performance} improve general serving, the unique combination of multi-stage pipeline dependencies and GPU resource heterogeneity in agentic Text-to-SQL serving is still largely unexplored.

To address these challenges, we introduce \sys, a novel framework that efficiently schedules and executes agentic Text-to-SQL workloads in heterogeneous GPU-serving environments while supporting multi-tenant queries with diverse SLOs. \sys is built around a \textit{two-level} scheduling architecture that directly responds to \underline{C1}-\underline{C3}:

\begin{itemize}
    \item To tackle \underline{C1}, \sys explicitly tracks workflow progress and dispatches dependent inference requests only when prerequisites complete. Parallelizable requests are dispatched concurrently to reduce stage completion time.

    \item To address \underline{C2}, \sys employs a workload-balanced dispatcher at the global coordinator, selecting a model instance by jointly considering (\underline{i}) the request's computational demand and (\underline{ii}) each instance's serving capacity and current load. This design ensures that heavy or latency-sensitive tasks run on powerful hardware, while lighter tasks utilize underutilized resources, achieving balanced utilization on a heterogeneous cluster.

    \item To satisfy \underline{C3}, each model instance maintains an urgency-guided local priority queue that \fred{reorders pending LLM requests at batch boundaries during continuous batching based on remaining budget and estimated execution time.} Per-query SLO budgets are continuously recomputed to propagate deadline pressure to downstreams.
\end{itemize}

Lastly, we summarize our contributions as follows:

\begin{itemize}
\item \textbf{Contribution 1.} We identify and formalize three core design principles for serving agentic Text-to-SQL workflows at scale: explicit multi-stage dependency management, heterogeneity-aware request allocation, and end-to-end SLO guarantees in multi-tenant settings.

\item \textbf{Contribution 2.} We propose \sys, a serving framework that combines the global workload-balanced dispatching policy with a local urgency-driven priority queue to enable SLO-aware execution on heterogeneous clusters.

\item \textbf{Contribution 3.} We conduct a comprehensive evaluation demonstrating \sys’s performance on agentic Text-to-SQL workloads. Across diverse traces and hardware configurations, \sys reduces P95 tail latency by $1.42{\sim}1.56\times$ and increases throughput by $1.49{\sim}1.81\times$ compared to state-of-the-art LLM serving systems.
\end{itemize}


\section{Preliminaries}
\label{sec:Preliminaries}

In this section, we first introduce the key concepts underlying our serving system design: \S\ref{sec:multistage} decomposes the Text-to-SQL workflow into its constituent stages, highlighting the sequential dependencies and multi-stage pattern that necessitate specialized scheduling for the workflow; \S\ref{sec:llmserving} analyzes limitations of the current scheduling and queuing policy in existing LLM serving systems, demonstrating why general-purpose schedulers fail to meet the end-to-end latency requirements of the agentic LLM-based Text-to-SQL workflow. 

\subsection{Agentic LLM-based Text-to-SQL Workflow}
\label{sec:multistage}
The agentic LLM-based Text-to-SQL workflow involves several key stages to transform a natural language query into an executable SQL statement. 
As shown in Figure~\ref{fig:t2qworkflow}, we summarize the key stages in the state-of-the-art agentic Text-to-SQL paradigm (mainly following CHESS \cite{talaei2024chess}) as below:

\begin{figure}
    \centering
    \includegraphics[width=0.6\linewidth]{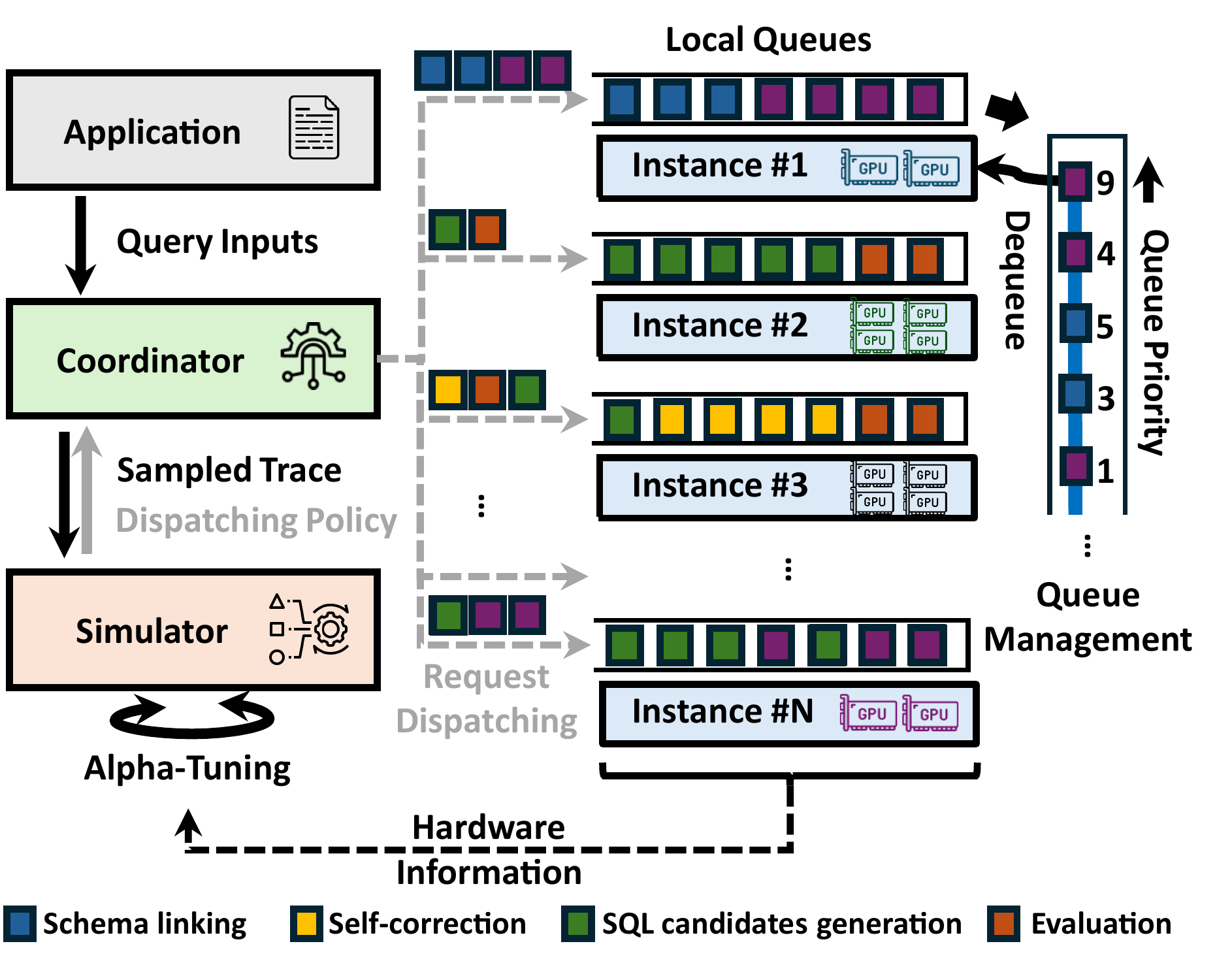}
    \caption{\sys system. Incoming LLM inference requests are dispatched by a global coordinator to model instances based on workload balance and task suitability. Each model instance manages its queue using an urgency-guided priority mechanism. Different color chunks represent an LLM inference request from a different stage in the workflow.}
    \label{fig:sys_arch}
    \vspace{-1.5em}
\end{figure}

\begin{itemize}
    \item \textbf{Schema linking}: 
    Given metadata and column descriptions for tables, the LLM identifies and aligns entities mentioned in the user's natural language query to relevant tables and columns within the database schema. This step is essential to accurately ground the user's query in the underlying database structure.

    \vspace{0.15em}
    \item \textbf{SQL candidates generation}: 
    Utilizing schema alignments from the previous step, the LLM generates a fixed number of candidate SQLs from the natural language query. Multiple LLM inference requests are executed concurrently, employing different prompts and illustrative examples, to produce a diverse set of candidates. This parallel approach aims to capture multiple interpretations of the user’s intent.

    \vspace{0.15em}
    \item \textbf{Self-correction}: 
    Candidate SQL queries are executed in the database, and any execution errors prompt iterative refinement throughout subsequent LLM invocations. The agent can iteratively refine queries up to a predefined limit (e.g., 10 iterations), systematically improving their correctness.

    \vspace{0.15em}
    \item \textbf{Evaluation}: 
    After self-correction removes execution errors, the LLM synthesizes a checklist-style set of “unit tests” in natural language (i.e., verifiable criteria derived from the user query). \fred{An LLM-as-a-judge scores each finalized SQL candidate against these criteria and selects the highest-scoring query. These tests verify both the semantic accuracy and functional correctness of the SQL candidates, ensuring alignment with the user's original intention.}

\end{itemize}

\subsection{Scheduling in LLM Serving}
\label{sec:llmserving}

Current LLM serving frameworks, such as vLLM~\cite{kwon2023efficient}, Text Generation Inference (TGI) from Hugging Face~\cite{huggingfaceTGI}, and TensorRT-LLM from NVIDIA~\cite{tensorRT}, employ various scheduling and queueing strategies optimized for general-purpose LLM serving. For example, vLLM utilizes continuous batching~\cite{yu2022orca} with a first-come-first-served (FCFS) policy, allowing new LLM inference requests to join ongoing batches during decoding. TGI groups incoming LLM inference requests based on prompt length and generation parameters to optimize batch formation. TensorRT-LLM implements an ``in-flight'' batching strategy, managing concurrent LLM inference requests through the scheduling policies implemented in Triton~\cite{nv_triton}, including FIFO and priority-based queuing. While effective for standard LLM tasks, these frameworks are not inherently designed to handle the complexities of multi-stage, dependency-aware agentic workflows like Text-to-SQL.
Advanced queuing methods, such as Queue Management for LLM serving (QLM)~\cite{patke2024QLM} and the Virtual Token Counter (VTC)~\cite{sheng2024fairness}, have been proposed to address specific performance and fairness requirements---QLM introduces priority scheduling to meet SLOs by prioritizing urgent LLM inference requests and employing techniques like preemption and state swapping; VTC ensures equitable resource allocation among users by tracking the number of tokens served and prioritizing those with lower consumption. However, these mechanisms still can not inherently account for the nuances of the Text-to-SQL pipeline, such as varying computational costs across different stages or the impact of query latency on downstream database performance.

\section{\sys}
\label{sec:sys}

This section first introduces three system requirement that Text-to-SQL serving systems should meet (\S\ref{sec:anatomy}), then demonstrates the framework overview  (\S\ref{sec:sysdesign}), and enumerates each core functionality (\S\ref{sec:impl}).

\subsection{Anatomy of Text-to-SQL Serving System}
\label{sec:anatomy}

To eliminate the limitations of current LLM serving systems serving Text-to-SQL workflows, we present three essential requirements for an agentic LLM-based Text-to-SQL system.

\begin{itemize}
\item \textbf{Explicit multi-stage dependency management.}
Agentic Text-to-SQL decomposes an end-to-end query into sequential and parallel stages (schema linking, candidate generation, self-correction, evaluation), each triggering one or more LLM inference requests. The system should explicitly track inter-stage dependencies, dispatching a stage only after prerequisites complete, while executing parallelizable requests concurrently to reduce stage completion time.

\vspace{0.15em}
\item \textbf{Heterogeneity-aware LLM inference request allocation.}
Production environments commonly feature hardware heterogeneity—mixed GPU deployments with varying computational capabilities, memory capacities, and performance characteristics due to incremental upgrades or cost considerations. Efficient serving therefore requires routing each LLM inference request to a suitable model instance by accounting for both the request’s compute demand (e.g., token lengths, memory footprint) and each instance’s serving capacity and current load~\cite{mei2025helix,jiang2024hexgen,jiang2025hexgen}.

\vspace{0.15em}
\item \textbf{End-to-end request SLO guarantees.}
Production-level Text-to-SQL serving systems must adhere to end-to-end SLOs to ensure consistent, predictable performance in multitenant environments with diverse workloads. Since a query spans multiple dependent LLM calls, the system should track per-query deadlines and prioritize requests based on remaining budget and expected execution time, adapting priorities as workflow progress consumes slack.
\end{itemize}


\subsection{\sys Overview}
\label{sec:sysdesign} 

To satisfy the requirements listed in \S\ref{sec:anatomy}, we propose \sys, a distributed system designed to efficiently serve multi-stage, LLM-based Text-to-SQL inference workloads in heterogeneous GPU clusters. Figure~\ref{fig:sys_arch} provides an overview of the system architecture. \sys's serving architecture centers on a centralized global coordinator and multiple LLM model instances over heterogeneous GPUs. The system handles multi-stage inference workflows under heterogeneous GPU deployments while meeting strict per-query SLO deadlines through intelligent dispatch and prioritization using a \textit{two-level} scheduling design: (\underline{i}) the global coordinator assigns each incoming LLM inference request to an appropriate model serving instance, considering the computational requirements of the request and the availability of each model instance; and \fred{each model instance maintains its own local priority queue, dynamically ordering pending inference requests by urgency. After dispatch, requests are processed at the selected instance, where pending requests can be continuously reordered as their urgency increases with waiting time. Correspondingly, the global coordinator focuses on making queue-aware decisions for new arrivals rather than migrating in-flight requests across instances, because transferring KV cache and batch state is costly and can negate latency gains.
}

\subsection{System Implementation}
\label{sec:impl}

We enumerate the core functionalities in \sys system here, the corresponding formulation and detailed solution of the scheduling algorithm are introduced in \S\ref{sec:algorithm}. 

\vspace{0.25em}
\noindent\textbf{Multi-stage dependency management.}
\sys treats each end-to-end Text-to-SQL query as a workflow of dependent LLM inference requests. The global coordinator tracks per-query stage completion and dispatches a request only when its prerequisites finish, ensuring correct execution order. For parallel stages (e.g., candidate generation), it dispatches ready requests concurrently across available instances to reduce stage latency. As stages complete, the coordinator updates the query’s remaining deadline, which tightens downstream budgets and increases their urgency in local queues.

\vspace{0.25em}
\noindent\textbf{LLM inference request allocation.}
\sys routes each request to a suitable model instance under heterogeneous GPUs. The global coordinator maintains empirical performance profiles to estimate a request’s execution time on each instance and tracks each instance’s backlog. It then dispatches the request by balancing (\underline{i}) estimated computation cost and (\underline{ii}) queueing delay, avoiding load-agnostic policies such as round-robin. \fred{If an instance becomes slow or unavailable, \sys relies on standard serving-layer timeout/retry semantics: the coordinator treats the attempt as failed and re-issues the request to another instance based on the same workload-balanced dispatching policy.} This routing improves utilization and tail latency by matching heavy requests to faster instances while filling residual capacity on slower ones.

\vspace{0.15em}
\noindent\textbf{Adaptive local priority queue scheduling.}
Each model instance maintains a local priority queue that selects requests at batch boundaries based on urgency. \sys derives per-request SLO budgets from the query’s remaining end-to-end deadline and estimated per-stage costs, and recomputes budgets as the workflow progresses. Requests that are closer to exhausting their budgets are prioritized, which reduces deadline misses under multi-tenant contention. We formalize budget allocation and the urgency metric in \S\ref{sec:PQ}.

\section{Scheduling Algorithm}
\label{sec:algorithm}

In this section, we introduce the mathematical formulation and solution of the global coordinator and local priority queue. We first formulate the scheduling problem as below:

\vspace{0.25em}
\noindent \textbf{Problem formulation.}
Consider a stream of Text-to-SQL queries $\{Q_1, Q_2, \ldots\}$ where $Q_i \sim \mathbb{P}_Q$. Each query $Q_i$ consists of LLM inference requests $\{q_{i,1},\ldots,q_{i,n_i}\}$ with an end-to-end SLO $T^{\text{SLO}}_i$. Given $N$ model instances $\mathbf{M}=\{m_1,\ldots,m_N\}$, request $q_{i,j}$ experiences instance-dependent service time $t_{i,j}^m$ (queueing + computation) if executed on $m$. We seek a dispatching policy $\phi$ that maps each request to an instance, $\phi(q_{i,j}) \in \mathbf{M}$, to maximize the probability of meeting end-to-end SLOs:
\begin{equation}
\arg\max_{\phi} \mathbb{P}\left(\sum_{j=1}^{n_i} t_{i,j}^{\phi(q_{i,j})} \le T^{\text{SLO}}_i \mid Q_i \sim \mathbb{P}_Q\right).
\end{equation}
Finding optimal policies under unknown arrival and service distributions is intractable---The inherent uncertainty in job characteristics necessitates dynamic decision-making without complete information, rendering the scheduling problem NP-hard~\cite{bhat2008introduction}. We therefore use a practical heuristic that combines global dispatching with local deadline-driven prioritization. Concretely, our design includes the following components:

\begin{itemize}
    \item \textbf{Global coordinator level (\S\ref{sec:WB}).}
    The coordinator performs workload-balanced dispatching, assigning each incoming LLM inference request to a model instance by considering both estimated execution time and current queueing delay.

    \vspace{0.15em}
    \item \textbf{Local priority queue level (\S\ref{sec:PQ}).}
    Each model instance maintains a priority queue and reorders pending LLM inference requests at batch boundaries using a deadline-aware urgency metric derived from per-request SLO budgets.

    \vspace{0.15em}
    \item \textbf{Simulator-guided tuning (\S\ref{sec:Simulator}).}
    A trace-driven simulator tunes the dispatching hyperparameter online to adapt to workload and hardware dynamics.
\end{itemize}

Our design ensures optimal exploitation of heterogeneous model instance serving capabilities and reliable compliance with strict per-query latency objectives in dynamic, multi-tenant deployments. Note that \sys will not affect the original Text-to-SQL results, where the scheduling does not modify model inputs or the inter-stage dependencies within the workflow. We summarize the notations we are going to use in this section in Table~\ref{tab:symbol} for reference. 

\input{symbol_table}

\subsection{Workload-Balanced Dispatching Policy} 
\label{sec:WB}

The workload-balanced dispatching policy in \sys assigns each LLM inference request to the most appropriate LLM serving model instance. This policy jointly considers (\underline{i}) the execution time of each instance for the incoming LLM inference request and (\underline{ii}) the current queueing time on each model instance. Additionally, a tunable hyperparameter $\alpha \in [0, 1]$ is introduced to dynamically balance the trade-off between these two factors. We discuss how to tune $\alpha$ in \S\ref{sec:Simulator}.

\vspace{0.25em}
\noindent \textbf{Formulate the inference computation cost.}
For each LLM inference request $q_{i,j}$, we first estimate its output length via a function $\hat{L}_{\text{out}}(q_{i,j})$ derived from its input length---our implementation is based on the prediction method introduced by Zheng et al.~\cite{zheng2023response}. Based on the estimation, we can get the predicted computational execution cost of $q_{i,j}$ on a particular model instance $m$, denoted ${t_{\text{comp}}}_{i,j}^m$, by the following equation:%
\begin{equation}
\label{eq:comp}
{t_{\text{comp}}}_{i,j}^m = {t_{\text{prefil}}}^m\left(L_{\text{in}}(q_{i,j})\right) + {t_{\text{decode}}}^m\left(\hat{L}_{\text{out}}(q_{i,j})\right)
\end{equation}%

\noindent where ${t_{\text{prefil}}}^m\left(L_{\text{in}}(q_{i,j})\right)$ and ${t_{\text{decode}}}^m\left(\hat{L}_{\text{out}}(q_{i,j})\right)$ denote the estimated execution time of the prefill and decoding phase on model instance $m$ based on number of input tokens and the estimated output tokens. 

\vspace{0.25em}
\noindent \textbf{Formulate the (maximal) queueing cost.}
The expected queuing time cost of $q_{i,j}$ on instance $m$, denoted by ${t_{\text{queue}}}_{i,j}^m$, is estimated by as the sum of the execution costs of all tasks currently in model instance $m$'s queue $\Theta^m$:
\begin{equation}
{t_{\text{queue}}}_{i,j}^m = \sum_{{q_{i',j'} \in \Theta^m}}  {{t_{\text{comp}}}_{i',j'}^m}
\end{equation}
Note that ${t_{\text{queue}}}_{i,j}^m$ captures the potentially longest time task $q_{i,j}$ could wait before execution begins if it is dispatched to model instance $m$.


\vspace{0.25em}
\noindent \textbf{Select the serving model instance.}
Given the estimation of inference computation time and queuing time, an ideal instance has a low estimation for both of them. However, a linear combination of these two factors is problematic---the execution time is relatively predictable given the LLM inference query, while the queuing time can be aggressively adjusted based on the urgency we implement within each local priority queue at each model instance. Thus, we define the following non-linear combination as the heuristic score:
\begin{equation}
\label{eq:score}
\text{Score}\left(q_{i,j},m\right) = (1 - \alpha) \cdot \frac{\beta}{{t_{\text{queue}}}_{i,j}^m} - \alpha \cdot {t_{\text{comp}}}_{i,j}^m
\end{equation}
\fred{For the LLM inference request $q_{i,j}$ and model instance $m$, $q_{i,j}$ will be dispatched to the instance with the highest score. Notice that there are two parameters in this heuristic score ($\alpha$ and $\beta$): $\alpha$ is tuned online, while $\beta$ is a deployment-specific normalization constant that makes the queue-sensitive term and the computation time term numerically comparable. Concretely, we have $\beta \triangleq C_{\text{ref}} \cdot Q_{\text{ref}}$, where $C_{\text{ref}}$ is a reference computation time and $Q_{\text{ref}}$ is a reference queuing time. With this definition, the queue-related term can be interpreted as a computation time equivalent reward: when $t_{\text{queue}}=Q_{\text{ref}}$, the reward equals $C_{\text{ref}}$; it increases when the target instance's queue is shorter than the reference level and decreases when it is longer. Since $C_{\text{ref}}$ and $Q_{\text{ref}}$ are primarily determined by stable deployment properties and change only when the deployment configuration materially changes, we calibrate $\beta$ once per deployment via few warmed-up queries and then keep it fixed. The weighting factor $\alpha$ determines the degree to which dispatching favors fast execution versus load balancing. When $\alpha = 1$, only execution speed is considered; when $\alpha = 0$, only queue-related signal matters. We determine an optimal $\alpha$ via sliding window-based monitoring and simulations (see \S\ref{sec:Simulator}).}


\subsection{Local Priority Queue} 
\label{sec:PQ}

\begin{figure}[t!]
    \centering
    \includegraphics[width=0.6\linewidth]{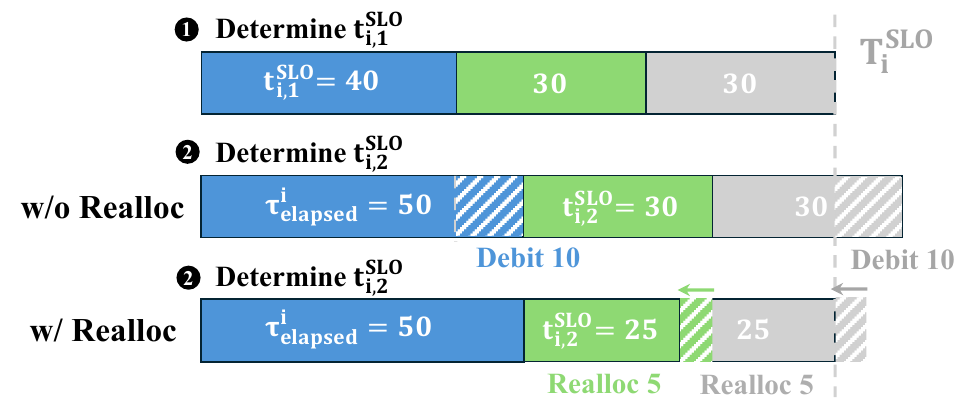}
    \caption{The SLO budget assigned to $q_{i,1}$ is $40$ms, but $q_{i,1}$ takes $50$ms to finish. The additional 10ms debit is recorded by $\tau^{i}_{\text{elapsed}}$ and reallocated to the following inference requests, resulting in a tighter SLO budget for $q_{i,2}$.}
    \label{fig:slo}
    \vspace{-1.5em}
\end{figure}

To meet the end-to-end SLO, the local priority queue includes two advanced mechanisms:

\vspace{0.25em}
\noindent \textbf{SLO budget allocation.} To determine the priority of each LLM inference request in the local queue, we allocate a per-request SLO budget $t^{\text{SLO}}_{i,j}$ for $q_{i,j}$ based on both execution cost and the remaining end-to-end deadline:
\begin{equation}
t^{\text{SLO}}_{i,j} = \left(T^{\text{SLO}}_i - \tau^{i}_{\text{elapsed}}\right) \cdot \frac{{\overline{t}_{\text{comp}}}_{{i,j}}}{\sum_{k=j}^{n_i} {\overline{t}_{\text{comp}}}_{i,k}}
\end{equation}
where $\tau^{i}_{\text{elapsed}}$ is the time elapsed since the arrival of $Q_i$ at the global coordinator, and ${\overline{t}_{\text{comp}}}_{i,k}$ is the estimated execution time of LLM inference $q_{i,j}$ averaged over all of the model instances, i.e., 
\begin{equation}
{\overline{t}_{\text{comp}}}_{i,j}=\frac{1}{N}\sum_{m\in\mathbf{M}} {t_{\text{comp}}}_{i,j}^m 
\end{equation}
This proportional allocation ensures that more time is budgeted for costlier downstream LLM inference requests. It's worth noting that \fred{although the workflow contains dynamic iterations, $n_i$ can be determined using configured bounds when available (e.g., a maximum refinement limit) together with the query's observed progress to estimate the remaining work. Additionally, unlike the preset end-to-end SLO $T^{\text{SLO}}_i$,} the per-request SLO budget $t^{\text{SLO}}_{i,j}$ is determined right before the inference request being sent. As shown in Figure~\ref{fig:slo}, $t^{\text{SLO}}_{i,j}$ is recomputed in real time as $\tau^{i}_{\text{elapsed}}$ is updated upon completion of $q_{i,j-1}$. This design ensures that Text-to-SQL queries behind schedule receive progressively tighter SLO constraints.

\vspace{0.25em}
\noindent \textbf{Urgency-based priority ranking.}
\fred{Our urgency metric is inspired by Least-Laxity-First (LLF) style deadline-driven scheduling~\cite{oh1998modified}.} For an LLM inference request $q_{i,j}$ dispatched to model instance $m$, we define the urgency $U_{i,j}$ of $q_{i,j}$ as the difference between its execution cost and the remaining SLO margin:
\begin{equation}
\label{eq:priority}
U_{i,j} = {t_{\text{comp}}}_{i,j}^m - \left(t^{\text{SLO}}_{i,j} - \tau_{i,j}\right)
\end{equation}
where ${t_{\text{comp}}}_{i,j}^m$ is the estimated execution cost of the task on instance $m$ as we defined in Equation~\ref{eq:comp}, and $\tau_{i,j}$ denotes the actual queuing delay tracked by the local priority queue since $q_{i,j}$ entered the local queue. Higher urgency indicates greater risk of SLO violation. \fred{The urgency score is time-aware and updated continuously as a request waits. As $\tau_{i,j}$ increases, the request's urgency increases, which prevents indefinite starvation even in the presence of competing high-urgency requests.} To handle estimation errors, any unexpected execution cost will be reflected in the per-request SLO budget $t^{\text{SLO}}_{i,j+1}$ of the subsequent inference request $q_{i,j+1}$.
Note that the priority is re-evaluated when a new batch is launched for inference, allowing urgent tasks to be admitted ahead of less critical ones. Whenever a request completes, the model instance $m$ always selects the LLM inference request with the highest urgency from the local queue:
\begin{equation}
q^* = \arg\max_{q_{i,j} \in \Theta^m} U_{i,j}
\end{equation}
where $\Theta^m$ is the current set of queued LLM inference requests at the model instance $m$. This adaptive strategy prioritizes requests at risk of missing their SLO and accounts for instance-specific execution characteristics.

\subsection{$\alpha$-Tuning Process}
\label{sec:Simulator}

The trade-off determined by $\alpha$ reflects an important trade-off in dispatching. Assigning each request to its most suitable model instance improves per-query efficiency, but can lead to uneven load distribution and queue buildup. On the other hand, balancing load evenly may assign queries to suboptimal instances, increasing latency. The hyperparameter $\alpha$ in the dispatching score function (Equation~\ref{eq:score}) allows the system to navigate this trade-off between execution cost and queuing cost adaptively. To tune $\alpha$ in response to real-time workload conditions, we implement a lightweight online parameter tuning process based on \textit{tail latency}, i.e., the delays experienced by the slowest requests in a system, typically the $95^{\text{th}}$ to $99^{\text{th}}$ percentile of response time~\cite{eddine2025tail}.

\vspace{0.25em}
\noindent \textbf{Initialization.}
During the cold-start of \sys system, the initial $\alpha^*$ is initialized to $0$, prioritizing queue length minimization. During the first short period of operation, \sys uses this policy to serve incoming Text-to-SQL queries. In parallel, the system collects execution trajectory, including arrival times, queue delays, and stage durations for each LLM inference request. Such information is essential for later simulation-guided auto-tuning.

\vspace{0.15em}
\noindent \textbf{Sliding window-based monitoring.}
After initialization, \sys assumes workload stationarity over short intervals and continues using the current $\alpha^*$. \fred{The system monitors tail latency in a sliding window. The window length is set to ensure a minimum number of completed end-to-end queries (e.g., $5$ samples) for a stable tail-latency estimate, and it adapts to workload changes such that heavier workloads typically result in longer windows.} At the end of each sliding window, it computes the $95^{\text{th}}$ percentile latency $\bar{T}_{\text{new}}$ for all queries served in that period and compares it against the baseline latency $\bar{T}_{\text{ref}}$ from the previous window.

\fred{While sudden workload shifts (e.g., peak bursts or anomalies) can cause performance degradations, we perform a one-sided two-sample $t$-test to determine whether latency has degraded significantly}:
\begin{equation}
H_0: \bar{T}_{\text{new}} = \bar{T}_{\text{ref}} \quad \text{vs.} \quad H_1: \bar{T}_{\text{new}} > \bar{T}_{\text{ref}}
\end{equation}
If the $p$-value falls below $0.01$, the null hypothesis is rejected, indicating statistically significant latency regression. This triggers a tuning procedure using the most recent trace. 

\vspace{0.15em}
\noindent \textbf{Simulation-guided auto-tuning.}
Auto-tuning is carried out using a trace-driven simulator that replays historical Text-to-SQL queries and evaluates performance under different $\alpha$ values on the fly. The optimal $\alpha^*$ is determined by minimizing the $95^{\text{th}}$ percentile latency (tail-aligned SLO objective):
\begin{equation}
\alpha^* = \arg\min_{\alpha \in [0,1]} T_{0.95}(\alpha)
\end{equation}
where $T_{0.95}(\alpha)$ denotes the simulated completion time that $95\%$ queries will be completed under parameter $\alpha$. The search iterates over $\alpha \in \{0.0, 0.1, 0.2, \ldots, 1.0\}$ in uniform steps.

\vspace{0.15em}
\noindent \textbf{Discussion of tuning overhead.}
The simulator executes entirely on the CPU and only incurs negligible overhead cost compared to the actual serving. In practice, tuned $\alpha^*$ values remain stable across adjacent time windows unless there are abrupt changes in workload patterns. This enables robust and low-overhead adaptation to evolving serving conditions, ensuring sustained low-latency performance.

\section{Evaluation}
\label{sec:evaluation}
In this section, we evaluate the design of \sys. We begin by asking the following questions to analyze the end-to-end performance of our framework, as well as each component's contribution towards the overall efficiency:

\begin{itemize}
    \item \textit{How does the end-to-end performance of \sys compare to that of state-of-the-art LLM inference serving systems in Text-to-SQL workflows?}

    \vspace{0.15em}
    \item \textit{How effective is each component during scheduling?}

    \vspace{0.15em}
    \item \textit{What are the benefits and cost of $\alpha$-tuning process?}
\end{itemize}

\noindent We state our experiment setup in \S\ref{sec:setup}, and present the experimental results to answer each question in \S\ref{sec:end2end}, \S\ref{sec:ablation}, and \S\ref{sec:alpha}, respectively.

\subsection{Experimental Setup} 
\label{sec:setup}

\noindent \textbf{Runtime.}
Each end-to-end Text-to-SQL query was processed following a state-of-the-art agentic framework, i.e., \textsc{CHESS}~\cite {talaei2024chess} that, at the time of this study, represents the leading approach in Text-to-SQL workflow. We perform evaluations in the following execution environments:

\begin{itemize}
\item  \underline{Hetero-1}: This setup consists of two types of GPUs, A100 and A6000, each supporting two model instances, with four GPUs allocated per instance for inference serving.
\vspace{0.15em}
\item \underline{Hetero-2}: This setup consists of three types of GPUs: A100, L40, and A6000. A100 GPUs are responsible for serving two model instances, while L40 and A6000 GPUs each are responsible for serving one model instance. Each model instance is allocated four GPUs for inference serving.
\vspace{0.15em}
\item \underline{Homo}: This setup consists solely of A100 GPUs, where the cluster hosts four model instances, each utilizing four GPUs for inference serving.
\end{itemize}

Note that due to the large number of model parameters, we employ a tensor parallelism degree of eight for serving all model instances using vLLM~\cite{kwon2023efficient} as the LLM inference engine, and all model instances adopt the continuous batching technique during inference.

\vspace{0.15em}
\noindent \textbf{Model and dataset.} The LLM inference requests following the agentic workflow in \textsc{CHESS}~\cite{talaei2024chess} are conducted using the \textsc{Llama3.1-70B} model for the main experiments. We follow prior work to generate three workload traces based on the development set from BIRD-bench{~\cite{li2023can}}, which is a cross-domain dataset designed specifically for Text-to-SQL evaluation. Our testing traces are sampled from queries related to \texttt{financial} and \texttt{formula1} databases, incorporating ``simple,'' ``moderate'', and ``challenging'' queries defined by BIRD ($14\%$, $72\%$, $14\%$ respectively). In particular, \texttt{Trace 1} samples queries purely from the \texttt{financial} database, \texttt{Trace 2} samples queries from the \texttt{formula1}, and \texttt{Trace 3} contains queries from both databases. Depending on the complexity of the query, the workflow may require between zero and ten rounds of revision to refine the SQL query. We count the number of LLM calls per Text-to-SQL query across three trials, observing an average of $19.4$, $20.7$, and $21.1$ calls for \texttt{Trace 1}, \texttt{Trace 2}, and \texttt{Trace 3}, respectively, with corresponding variances of $12.1$, $14.7$, and $16.3$. The variances reflect the different complexity of sampled queries among traces, which widely exists in real-world Text-to-SQL workloads. \fred{To emulate the stochastic arrival pattern of users' Text-to-SQL queries, we send queries using a Poisson process with arrival rates of $0.5$ query per second and $1.0$ query per second (QPS), where each end-to-end query triggers around 20 LLM inference requests on average.} This modeling approach captures the inherent randomness and non-stationarity in user interactions and aligns with methodologies employed in prior studies~\cite{wang2024revisiting}. \fred{To further stress the system under heavier workload, we additionally evaluate a $30$~QPS setting on \texttt{Trace 3} as a stress test. Beyond stationary arrivals, we also construct a fluctuating workload with non-stationary arrival rates ($5$ QPS to $20$ QPS to $5$ QPS) to examine how our method adapts to abrupt changes in traffic intensity. We also demonstrate generalizability by employing an alternative agentic workflow, \textsc{MAC-SQL}~\cite{wang-etal-2025-mac}, with the \textsc{Qwen3-30B} model~\cite{qwen3}. For this evaluation, we sampled a \texttt{Trace 4} from Spider1.0~\cite{yu-etal-2018-spider}, containing 9 new domain databases including \texttt{dog kennels}, \texttt{cars}, etc. Finally, to evaluate robustness under hardware sharing, we derive \texttt{Trace 5} and \texttt{Trace 6} by augmenting \texttt{Trace 3} with $25\%$ and $50\%$ independent single-shot LLM inference requests, each with their own SLOs.}


\vspace{0.15em}
\noindent \textbf{Baseline.} To evaluate the end-to-end performance under three deployment setups: \underline{Hetero-1}, \underline{Hetero-2}, and \underline{Homo}, we compare \sys with five strong baselines:

\begin{itemize}
\item \textbf{vLLM}: We compare \sys against vanilla vLLM, which is a widely adopted inference serving system that uses first-come-first-served (FCFS) and continuous batching to manage local queues. We dispatch the LLM inference requests to model instances based on the round-robin strategy. This naive approach is commonly utilized in existing inference serving systems~\cite{down2006multi,grosof2019load}.

\vspace{0.15em}
\item \textbf{VTC}: We include the existing scheduling algorithm Virtual Token Counter (VTC)~\cite{sheng2024fairness} under all three traces. VTC defines fairness based on a cost function accounting for input/output token counts served for different users.
To leverage this design, we view one particular end-to-end Text-to-SQL query as a user in VTC, since each end-to-end Text-to-SQL query is composed of multiple inference requests sent in order. In such a way, VTC can balance the serving for various end-to-end Text-to-SQL queries.

\vspace{0.15em}
\item \textbf{QLM}: We also compare \sys with a SLO-oriented queue management system QLM~\cite{patke2024queue}, which groups inference requests based on their SLOs and uses a request waiting time estimator to estimate the waiting time of each request group in the local queue. If it estimates that the waiting time is too long to violate the group's SLO, all the request groups in the local queue will be reordered to prioritize the group with the least remaining time. Within each request group, inference requests are scheduled according to FCFS.

\vspace{0.15em}
\item \fred{\textbf{LLF}: We include a Least Laxity First (LLF)~\cite{oh1998modified} baseline that treats each Text-to-SQL query as a ``job'' with an end-to-end deadline. At runtime, for each stage we schedule inference requests by the job's laxity (i.e., deadline minus elapsed time and remaining processing time), without performing \sys's per-request budget recomputation.

\vspace{0.15em}
\item \textbf{Ray}: We compare \sys with Ray Serve~\cite{rayserve-docs}, a widely-used general-purpose online serving framework that can express our agentic workflow DAG by chaining deployments. Ray Serve routes requests using power-of-two-choices (Po2C)~\cite{azar1999balanced}; however, it does not incorporate cross-stage, end-to-end deadline-aware scheduling strategies.}

\end{itemize}

Additionally, we include two variants of \sys in the ablation study to assess the impact of our workload-balanced dispatching (WB) and local priority queue (PQ) independently of other enhancements in \sys:

\begin{itemize}
\item \textbf{RR+PQ}: We consider a truncated version of \sys, which replaces the WB mechanism in our global coordinator with a simple round-robin (RR) dispatcher combined with our PQ implementation on each model instance. 

\vspace{0.15em}
\item \textbf{WB+FCFS}: We also include another ablated version of \sys, which reserves the WB mechanism in our global coordinator, while utilizing simple FCFS scheduling on each model instance. 
\end{itemize}

\begin{figure*}[t!]
    \centering
    \includegraphics[width=\linewidth]{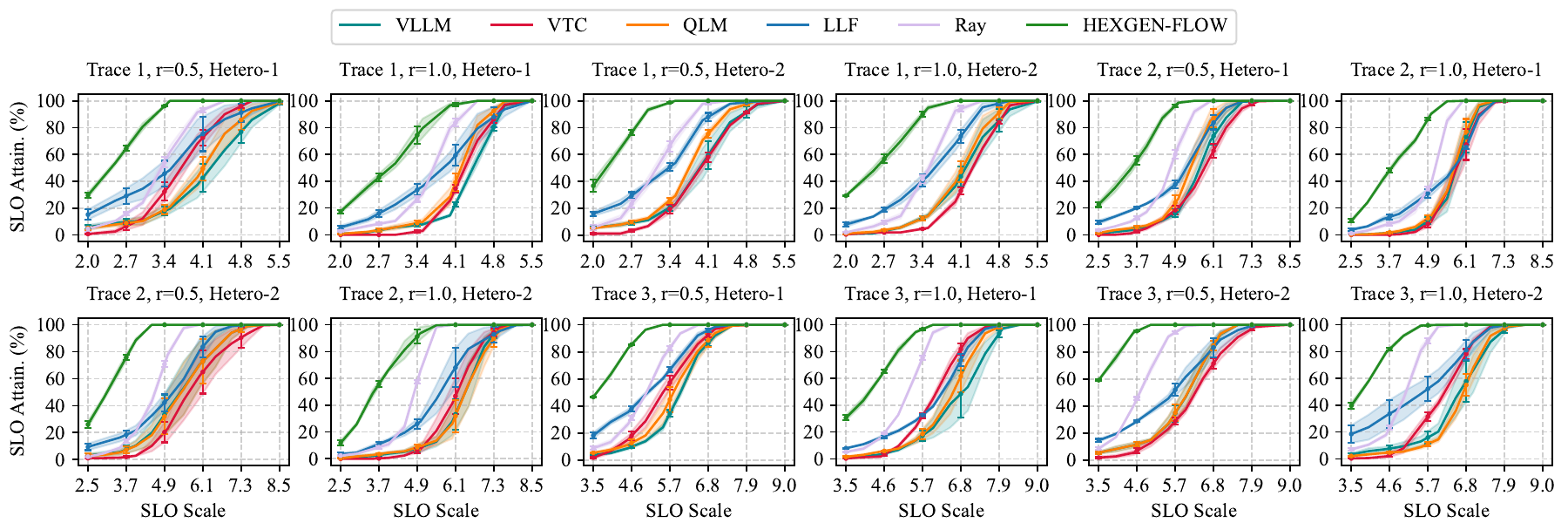}
    \caption{End-to-end SLO attainment comparison. $r$ represents the end-to-end queries arrival rate in queries per second.}
    \label{fig:exp1}
    \vspace{-0.5em}
\end{figure*}

\vspace{0.15em}
\noindent \textbf{Evaluation metrics.}
Following the standard setup of existing LLM serving frameworks~\cite{li2023alpaserve,zhong2024distserve}, we evaluate system performance based on SLO attainment and system throughput, where users specify the SLO. For evaluation purposes, we first collect the execution time of each end-to-end Text-to-SQL query in an exclusive environment. During our evaluation, we set the SLO by multiplying the exclusive execution time with various scales (i.e., SLO scale in Figure~\ref{fig:exp1}) to assess performance under different levels of operational stringency. Particularly, we are interested in understanding the performance under the setup with minimum SLO scale at which the system achieves $95\%$ SLO attainment (i.e., P95 tail latency). We define the throughput metric as the number of end-to-end Text-to-SQL queries completed per second. We also conduct additional experiments in the multi-tenant scenario where there are two distinct SLOs corresponding to two users.

\begin{figure*}[t!]
    \centering
    \includegraphics[width=\linewidth]{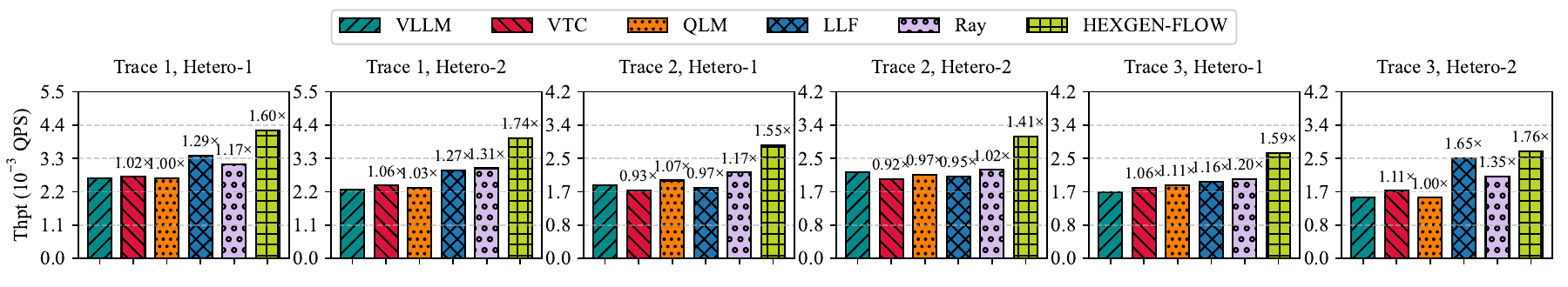}  
    \vspace{-2.0em}
    \caption{End-to-end system throughput comparison.}
    \vspace{-0.5em}
    \label{fig:thpt}
\end{figure*}

\begin{figure*}[t!]
    \centering
    \includegraphics[width=\linewidth]{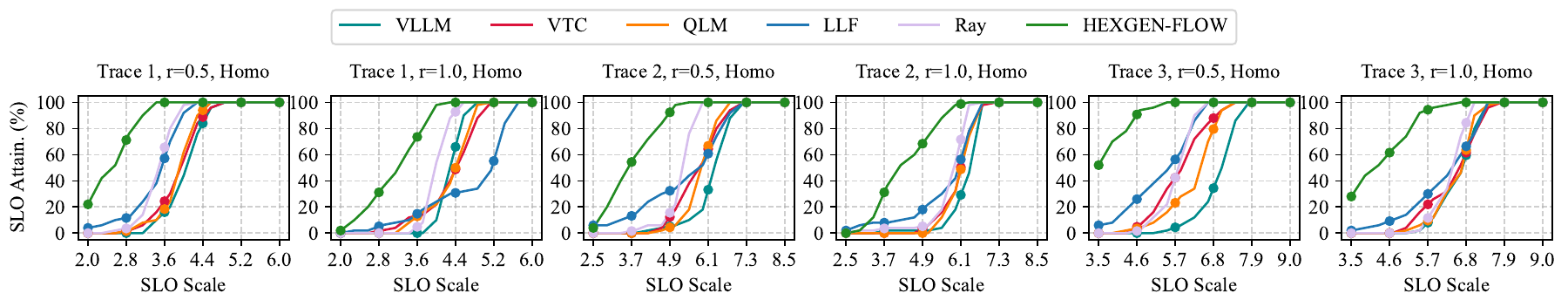}
    \vspace{-2.0em}
    \caption{End-to-end experiments of \sys under homogeneous settings.}
    \vspace{-0.75em}
    \label{fig:homo}
\end{figure*}

\subsection{End-to-End Performance Boost}
\label{sec:end2end}

\textbf{End-to-end system performance.} We evaluate the SLO attainment and throughput of \sys across multiple traces, system loads, and heterogeneous hardware configurations, comparing its performance against five baseline systems: vLLM, VTC, QLM, LLF, and Ray. We conduct multiple runs with random seeds $0, 1, 2$, and include error bars in the plot. Figure~\ref{fig:exp1} and Figure~\ref{fig:thpt} demonstrate \sys consistently outperforms all baseline systems, concretely:

\begin{itemize}
\item Compared with vLLM, \sys achieves P95 tail latency reductions ranging from $19.2\%$ to $56.2\%$, along with throughput improvements ranging from $49.1\%$ to $81.0\%$. For example, when evaluated on \texttt{Trace 1} with \underline{Hetero-1} under $0.5$ QPS system load, vLLM requires $5.4$ SLO scales to achieve $95\%$ attainment, whereas \sys requires only $3.5$ SLO scales. These results demonstrate that vLLM with naive request dispatching and scheduling methods (FCFS and round robin) fail to provide satisfactory serving performance for complex Text-to-SQL workloads.

\vspace{0.15em}
\item Compared with VTC, \sys reduces P95 tail latency by $19.2\%{\sim}53.8\%$ and increases throughput by $10.3\%{\sim}39.8\%$. Although VTC provides balanced resource allocation across different Text-to-SQL queries through its fairness-based scheduling mechanism, it fails to account for the heterogeneous query requirements specific to Text-to-SQL workflows, thereby leading to suboptimal system performance. For instance, when evaluated on \texttt{Trace 2} with \underline{Hetero-2} under $0.5$ QPS system load, VTC requires $7.9$ SLO scales to achieve $95\%$ attainment, while \sys requires only $5.2$.

\vspace{0.15em}
\item Compared with QLM, \sys achieves P95 tail latency reductions between $14.0\%$ and $39.6\%$, and throughput gains between $17.5\%$ and $36.6\%$. Although QLM utilizes an estimation-based queue management approach that prioritizes urgent requests to maximize SLO attainment, it applies uniform treatment to all queries without accounting for the diverse computational requirements inherent in Text-to-SQL workflows. This limitation compromises performance optimization. For instance, when evaluated on \texttt{Trace 3} with \underline{Hetero-2} under $1.0$ QPS system load, QLM requires $7.8$ SLO scales to achieve $95\%$ attainment, while \sys requires only $6.1$.

\vspace{0.15em}
\item \fred{Compared with LLF, \sys achieves P95 tail latency reductions by $12.4\%{\sim}39.6\%$, and improves throughput by $6.7\%{\sim}59.8\%$. Although LLF prioritizes ready stage calls by end-to-end job laxity, it does not perform per-request budget recomputation across stages; as a result, it may over- or under-allocate slack to intermediate calls, causing avoidable SLO violations and reduced throughput. For example, on \texttt{Trace 3} with \underline{Hetero-2} under $1.0$ QPS, LLF needs roughly $7.6$ SLO scales to reach $95\%$ attainment, while \sys reaches $95\%$ at around $5.2$.

\vspace{0.15em}
\item Compared with Ray, \sys reduces P95 tail latency by $8.6\%{\sim}23.8\%$ and delivers $30.4\%{\sim}38.2\%$ higher throughput. Ray Serve's routing strategy neglects the hardware heterogeneity, and its lack of cross-stage deadline-aware scheduling limits the performance. For instance, on \texttt{Trace 3} with \underline{Hetero-2} under $1.0$ QPS, Ray Serve requires approximately $5.9$ SLO scales to reach $95\%$ attainment, whereas \sys reaches $95\%$ at around $5.2$.}
\end{itemize}

\begin{figure*}[t!]
    \centering
    \includegraphics[width=\linewidth]{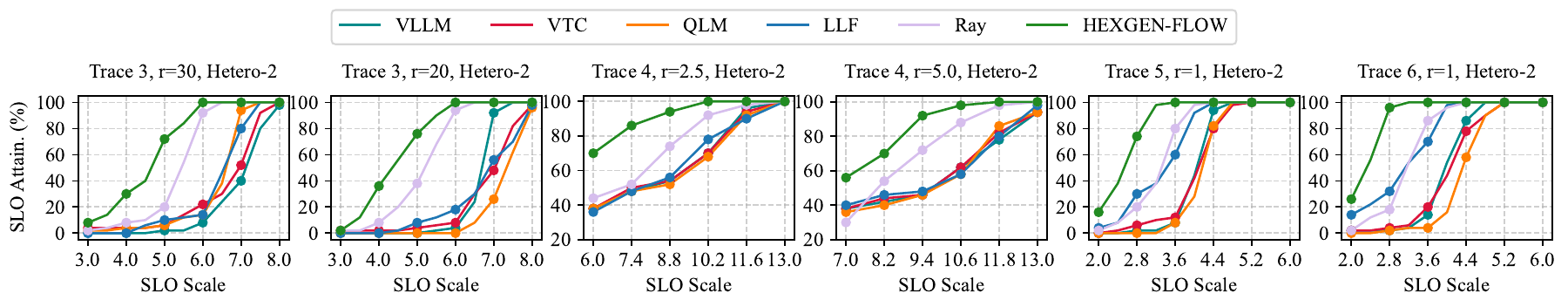}
    \vspace{-1.5em}
    \caption{\fred{End-to-end SLO attainment comparison. \textbf{\underline{Left two:}} results under a higher QPS and a workload with a sudden shift. \textbf{\underline{Middle two:}} results on the Spider trace. \textbf{\underline{Right two:}} results under 25\% and 50\% competing workloads. $r$ represents the end-to-end queries arrival rate in queries per second.}}
    \vspace{-1.0em}
    \label{fig:additional}
\end{figure*}

\vspace{0.15em}
\noindent \textbf{Homogeneous environment evaluation.} We also evaluate the SLO attainment of \sys in homogeneous environments across different traces and system loads. As shown in Figure~\ref{fig:homo}, \sys consistently outperforms all baselines in homogeneous configurations. Specifically, across all evaluated scenarios, \sys achieves up to $32.6\%$, $31.4\%$, $28.6\%$, $47.3\%$, and $19.2\%$ lower P95 tail latency compared to vLLM, VTC, QLM, LLF, and Ray respectively. When evaluated on \texttt{Trace 3} under $1.0$ QPS system load, existing baseline systems require a minimum of $7.2$ SLO scales to achieve $95\%$ attainment, whereas \sys requires only $5.9$ SLO scales. These results demonstrate that \sys delivers outstanding system performance even in homogeneous environments, further validating the effectiveness of our approach across diverse hardware configurations.

\vspace{0.15em}
\noindent \fred{\textbf{Evaluation under high and non-stationary arrival rates.} As shown in \textbf{\underline{left two}} of Figure~\ref{fig:additional}, under both the heavy ($30$ QPS throughout) and fluctuating ($20$ QPS during peak) workloads, \sys maintains strong tail-latency control. In particular, \sys reduces the P95 tail latency by $5.2\%{\sim}23.3\%$ compared to all the baselines. Moreover, in the fluctuating workload, $\alpha$-tuning is triggered twice, matching the two workload shifts and indicating timely adaptation.}

\vspace{0.15em}
\noindent \textbf{Evaluation on alternative workflows and models.} We conduct supplementary experiments on \textsc{MAC-SQL} using the \textsc{Qwen3-30B} model. The results demonstrate that our approach maintains consistent improvements across diverse models and tasks, highlighting its potential applicability to other agentic workflows as well. From \textbf{\underline{middle two}} of Figure~\ref{fig:additional}, \sys achieves P95 tail latency reductions ranging from $34.0\%$ to $38.3\%$ compared to baseline approaches.

\vspace{0.15em}
\noindent \fred{\textbf{Evaluation in competing workloads scenarios.} The \textbf{\underline{right two}} of Figure~\ref{fig:additional} further evaluate robustness under a broader range of non-Text-to-SQL competing workloads. Across $25\%$ and $50\%$ competing-load mixes, \sys reduces P95 tail latency by approximately $18.2\%$ to $36.4\%$ compared to baseline approaches, while sustaining higher SLO attainment across SLO scales.}

\begin{figure}
    \centering
    \includegraphics[width=0.5\linewidth]{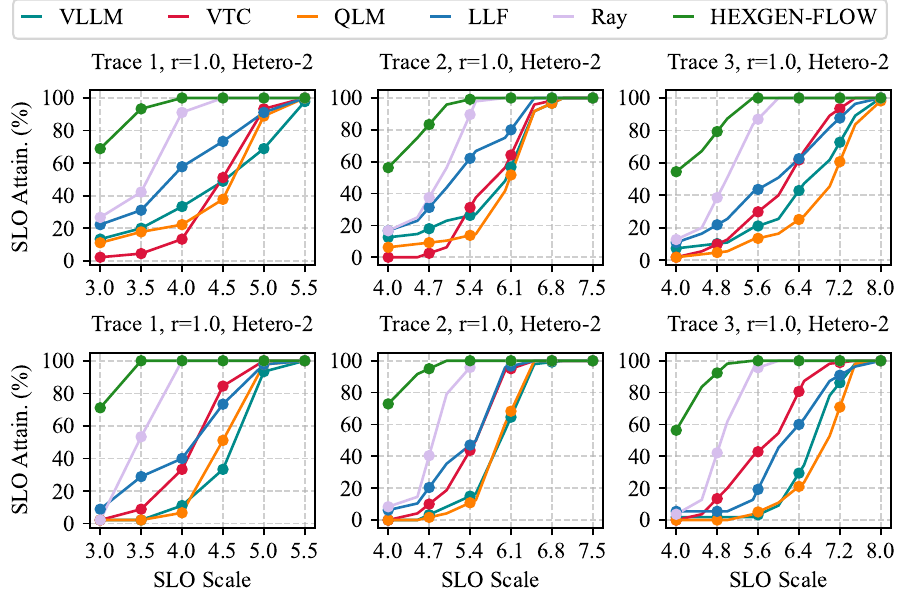}
    \caption{End-to-end SLO attainment comparison for Text-2-SQL queries from two users. The first row corresponds to user 1 and the second row corresponds to user 2. $r$ represents the end-to-end queries arrival rate in queries per second.}
    \vspace{-1.0em}
    \label{fig:tenant}
\end{figure}

\vspace{0.15em}
\noindent \textbf{Evaluation in multi-tenant scenarios.} In our design, each LLM inference request's urgency can adapt to various SLOs, allowing our algorithm to integrate seamlessly into multi-tenant scenarios. To evaluate \sys in scenarios involving multiple users with heterogeneous SLO requirements, we conduct additional experiments in a setting where two users with distinct SLOs coexist. The results in Figure \ref{fig:tenant} demonstrate that \sys reduces P95 tail latency by $27.9\%{\sim}35.7\%$ for users with different SLO requirements compared to baseline approaches. \fred{Additionally, based on per-tenant SLO attainment rates, we compute Jain's fairness index~\cite{jain1998fairness} to be 0.98 across three traces.}

\vspace{0.15em}

\begin{figure*}
    \centering
    \includegraphics[width=\linewidth]{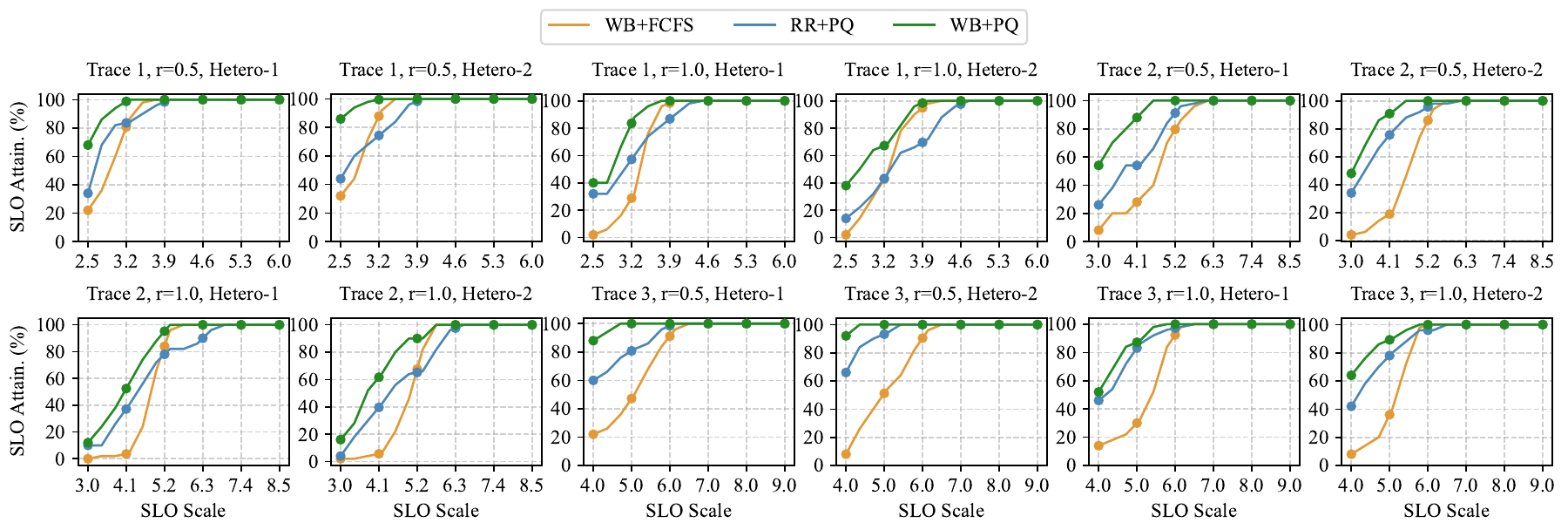}
    \caption{Ablation study of \sys's scheduling components, comparing end-to-end SLO attainment rates across: (1) round-robin dispatching + local priority queue, (2) workload-balanced dispatching + First Come First Serve, and (3) workload-balanced dispatching + local priority queue.}
    \vspace{-0.5em}
    \label{fig:exp2}
\end{figure*}

\subsection{Ablation Study: Effectiveness of Scheduling} 
\label{sec:ablation}

To assess the individual contributions of our WB and PQ optimizations, we conduct a systematic assessment by replacing each optimization with a naive baseline implementation (RR or FCFS), and evaluate the resulting SLO attainment. This evaluation spans multiple traces, system loads, and heterogeneous hardware configurations. Note that WB+PQ represents the complete implementation of \sys. As shown in Figure~\ref{fig:exp2}, the results demonstrate the distinct performance contributions of each optimization component:

\begin{itemize}
  \item \textbf{Effectiveness of WB.} We compare WB+PQ against RR+PQ to evaluate the effectiveness of WB. Experimental results demonstrate that WB consistently outperforms RR, achieving $10.5\%$ to $34.8\%$ lower P95 tail latency. For instance, when evaluated on \texttt{Trace 3} with \underline{Hetero-1} under $0.5$ QPS system load, RR requires $5.8$ SLO scales to achieve $95\%$ attainment, while WB requires only $4.3$ SLO scales, reducing the tail latency by roughly $35\%$. These results validate the effectiveness of WB in balancing workloads and resource utilization.

 \vspace{0.15em}
 \item \textbf{Effectiveness of PQ.} We compare WB+PQ against WB+FCFS to evaluate the effectiveness of PQ. Results demonstrate that PQ significantly outperforms FCFS across all evaluated scenarios, achieving up to $48\%$ lower P95 tail latency. For instance, when evaluated on \texttt{Trace 2} with \underline{Hetero-2} under $0.5$ QPS system load, FCFS requires $5.8$ SLO scales to achieve $95\%$ attainment, while PQ requires only $4.6$. These results further validate the design of PQ, and demonstrate the importance of urgency-based request prioritization in serving Text-to-SQL workloads.
\end{itemize}

\input{table1.tex}

\vspace{0.15em}
\noindent
\textbf{Case study of WB impact on request dispatching.} To demonstrate how WB improves request dispatching, we analyze task allocation across model instances with and without WB on \texttt{Trace 3} with \underline{Hetero-2} under $0.5$ QPS system load. As shown in Table~\ref{tab:optimization-results}, unlike RR which distributes tasks evenly across GPU instances, WB strategically allocates computationally intensive stages to more powerful instances (e.g., A100) while assigning lighter stages to less capable ones (e.g., L40). For example, stage 2 and stage 1 requests are evaluated as computationally intensive and light, respectively. Under WB, A100 instances collectively handle $57.5\%$ of stage 2 requests compared to only $18.1\%$ under RR, while the L40 instance processes $3.1\%$ of stage 2 requests versus $9.6\%$ under RR. Conversely, WB allocates $32.3\%$ of stage 1 requests to the L40 instance compared to $26.8\%$ under RR, while allocating only $1.4\%$ of stage 1 requests to A100 instances compared to $43.9\%$ under RR. As a result, WB ensures optimal resource utilization by matching computational requirements with hardware capabilities, thereby enhancing overall performance.

\vspace{0.25em}
\noindent 
\textbf{Case study of PQ impact on request processing.} To demonstrate how PQ optimizes request processing within local queues, we examine a model instance's local queue state before processing the next LLM call, as shown in Table~\ref{tab:waiting-queue}. Under FCFS, the system will process request 1, which arrived earliest at $22.4$ seconds. However, this approach disregards the varying urgency levels of requests, potentially leading to SLO violations when early-arriving requests with loose SLO budgets are prioritized over later-arriving requests with stricter SLO requirements. In contrast, PQ prioritizes requests based on urgency values instead of arrival times. In this scenario, PQ will select request 6 to process next, which has the highest urgency value of $26.9$ despite arriving at $64.4$ seconds. This intelligent prioritization ensures the most time-critical requests are processed first, thereby maximizing the likelihood of meeting SLO requirements across requests in the queue.

\input{table2.tex}

\begin{figure*}[t!]
    \centering
    \includegraphics[width=\linewidth]{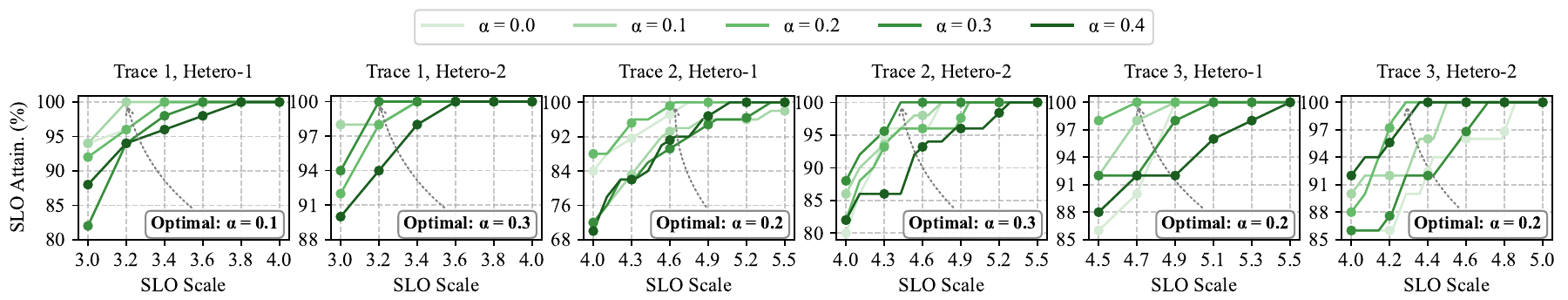}
    \caption{Performance of \sys under various $\alpha$ settings.}
    \vspace{-0.5em}
    \label{fig:exp3}
\end{figure*}

\subsection{Empirical Analysis of $\alpha$-Tuning} 
\label{sec:alpha}

We conduct experiments to estimate how dynamic tuning of the $\alpha$ parameter in \sys’s scheduling policy could accommodate hardware heterogeneity and workload variability. Through experiments across all three workloads (\texttt{Traces 1,2,3} and two heterogeneous GPU deployments (\underline{Hetero-1} and \underline{Hetero-2}), we evaluate: (\underline{i}) the impact of different $\alpha$ values on $95\%$ SLO attainment, and (\underline{ii}) the feasibility of simulation-based tuning during live serving. All experiments are conducted under a fixed query arrival rate of $0.5$ QPS, while varying $\alpha$ from $0$ (pure workload balancing) to $0.5$ (balanced weighting).

\vspace{0.10em}
\noindent \textbf{Effectiveness of $\alpha$-tuning.} 
Figure~\ref{fig:exp3} shows how different $\alpha$ values affect $95\%$ SLO attainment across traces and deployment configurations. Concretely, our plots focused on the $[0, 0.5]$ interval for readability purposes, as the optimum lies within that range. For \texttt{Trace 1}, $\alpha=0.1$ yields the best performance under \underline{Hetero-1}, while $\alpha=0.3$ performs optically under \underline{Hetero-2}. Similarly, the results in \texttt{Trace 2} indicate optimal values when $\alpha=0.2$ for \underline{Hetero-1}, and $\alpha=0.3$ for \underline{Hetero-2}. In \texttt{Trace 3}, $\alpha=0.2$ is optimal for \underline{Hetero-1}, while $\alpha=0.4$ performs best under \underline{Hetero-2}, improving $95\%$ SLO attainment latency by $14\%$ over $\alpha=0$. These results suggest that while workload balancing is a dominant factor, incorporating task suitability (via non-zero $\alpha$) consistently improves scheduling effectiveness. Moreover, the optimal $\alpha$ varies depending on both hardware configuration and workload composition, indicating that adaptive tuning is essential.

\input{table3.tex}

\vspace{0.10em}
\noindent \fred{\textbf{Overhead of $\alpha$-tuning.} We also measure the overhead of simulation-based $\alpha$ tuning. Under $0.5$ and $1.0$ QPS, each sliding window lasts for $100$ seconds. As shown in Table~\ref{tab:overhead}, through parallelized simulations, it takes only $1.9{\sim}3.1$ seconds per window across different settings. Moreover, tuning is triggered only when a performance degradation is detected in a window; the trigger rate ranged from $28.6\%{\sim}40.0\%$ across traces. Under stress tests with $20$ and $30$ QPS load, the window lasts for $300$ seconds, while the tuning takes $3.9{\sim}6.6$ seconds per window, and only $28.6{\sim}33.3\%$ of windows are triggered. Notice this overhead is minor relative to typical workload drift timescales, which span hours~\cite{zhong2024distserve}, making periodic tuning feasible in practice.
}

\vspace{-0.25em}

\section{Related Work}
\label{sec:related}

In this section, we provide a brief summary about the recent advances in LLM-based Text-to-SQL approaches in \S\ref{sec:related_Text-to-SQL} and then discuss some serving system design and implementation for LLM inference requests in \S\ref{sec:related_llm_serving}.

\subsection{LLM Advances for Text-to-SQL}
\label{sec:related_Text-to-SQL}

LLMs have emerged as a revolutionary paradigm for the Text-to-SQL task~\cite{katsogiannis2023survey,gao2024text}, where the core technique lies in effective SQL generation and schema linking.

\vspace{0.25em}
\noindent\textbf{SQL generation.}
Some pioneer studies focus on better \textit{prompting strategies and LLM adaptations} to boost Text-to-SQL performance. Gu et al.~\cite{gu2023few} propose a structure- and content-based prompt learning method to enhance few-shot Text-to-SQL translation, while Li et al.~\cite{li2024codes} build an open-source LLM specialized for SQL generation. Other approaches fine-tune or constrain LLM outputs to improve accuracy: Ren et al.~\cite{ren2024purple} propose Purple, which refines a general LLM to make it a more effective SQL writer. 
Beyond simple LLM adaptations, recent \textit{agentic approaches} leverage multiple LLM inference requests to collaboratively accomplish the Text-to-SQL tasks. Fan et al.~\cite{fan2024combining} combine a small language model with a large one to handle different sub-tasks of Text-to-SQL, thereby improving zero-shot robustness. Similarly, Pourreza et al.~\cite{pourreza2024dts} explore task decomposition across models: DTS-SQL breaks the problem into stages handled by smaller LLMs sequentially, and their DIN-SQL approach has an LLM refine its own output through iterative self-correction~\cite{pourreza2023din} in the prompt. Another line of research enhances the reasoning process of LLMs to produce correct SQL. One strategy is to incorporate intermediate steps or a reasoning framework during inference. Zhang et al.~\cite{zhang2024reactable} apply the ReAct paradigm~\cite{yaoreact} to table question answering, which encourages the LLM to generate and reason with intermediate actions (e.g., decomposition or calculations) before finalizing the SQL query. 
Techniques have also been explored to optimize the context given to the LLM: Talaei et al.~\cite{talaei2024chess} present CHESS, which harnesses contextual information efficiently to guide the LLM's SQL synthesis without increasing model size. 

\vspace{0.15em}
\noindent\textbf{Schema linking.} Integrating database schema knowledge and domain-specific information into LLM-driven Text-to-SQL is another important procedure. Eyal et al.~\cite{eyal2023semantic} decompose the natural language question and the corresponding SQL query into semantic sub-units, improving the model's understanding of how question clauses align with schema elements. Dou et al.~\cite{dou2022towards} incorporate external knowledge (e.g., business rules or formulas) into the parsing process to handle queries that require facts beyond the database content. Several works specifically target schema linking challenges in the age of LLMs. Liu et al.~\cite{liu2025solid} propose Solid-SQL, which uses enhanced in-context examples to make an LLM more robust at matching question terms to the schema during generation. 

\subsection{LLM Inference Request Scheduling}
\label{sec:related_llm_serving}

Efficient scheduling of LLM inference requests is crucial in modern AI infrastructure, essential to meeting latency and throughput requirements, particularly under varying system constraints~\cite{fuefficient,agrawal2024taming,gao2025apt,ao2025optimizing,jain2025performance}.
In environments with \textit{consistent hardware and model setups}, scheduling techniques focus on optimizing latency and throughput. Patke et al.~\cite{patke2024queue} introduce QLM, a system that estimates request waiting times to prioritize tasks, ensuring that SLOs are met under load conditions. Gong et al.~\cite{gong2025past} propose the future scheduler, which uses historical workload patterns and predictive modeling to make informed scheduling decisions, fulfilling SLA guarantees. 
Fu et al.~\cite{fuefficient} frame LLM scheduling as a learning-to-rank problem, training models to order queued requests to optimize end-to-end latency and throughput, outperforming traditional heuristics. Agrawal et al.~\cite{agrawal2024taming} present Sarathi-Serve, a system that adjusts batching and resource allocation to balance throughput and latency, particularly effective for requests of high priority versus low priority.
In setups with varying hardware capabilities and model types, recently proposed scheduling strategies adapt to resource \textit{heterogeneity}: Wan et al.~\cite{borui2025efficient} develop BROS, a system that differentiates between real-time and best-effort LLM queries, ensuring interactive queries are prioritized without compromising background batch processing. 
Sun et al.~\cite{sun2024llumnix} introduce Llumnix, a dynamic scheduling system that adjusts resource allocation for LLM serving in real time as query load patterns change, demonstrating benefits in single-model scenarios. Very recently, Fang et al.~\cite{fang2025improving} investigate the efficiency of multi-LLM application pipelines in an offline setting, using sampling and simulation to optimize inference plans for workflows involving multiple LLMs or sequential model calls.
Despite these efforts, there is a notable gap in scheduling strategies that coordinate end-to-end pipelines with multiple dependent LLM request serving stages. Our approach builds upon existing insights, such as urgency-aware prioritization and hardware-sensitive allocation, but effectively extends them to manage complex, multi-stage Text-to-SQL workflows under strict latency requirements and system heterogeneity with a significant performance boost.

\vspace{-0.5em}
\section{Conclusion}
\label{sec:conclusion}
In this paper, we introduced \sys, a novel inference-serving framework specifically designed to efficiently schedule and execute multi-stage agentic Text-to-SQL workloads in heterogeneous GPU clusters handling multi-tenant requests. By systematically analyzing the unique characteristics of these workloads-including stage dependencies, task variability, resource heterogeneity, and stringent latency constraints---we formulated a hierarchical scheduling solution composed of a global workload-balanced dispatcher and a local urgency-driven queue in each LLM serving model instances. This \textit{two-level} scheduling strategy significantly improves resource utilization and effectively mitigates deadline violations. Comprehensive experiments demonstrated that \sys consistently outperforms the existing state-of-the-art LLM serving framework, achieving substantial reductions in query latency and notable improvements in system throughput under realistic production workloads---\sys reduces P95 tail latency by $1.42{\sim}1.56\times$ and increases throughput by $1.49{\sim}1.81\times$ compared to state-of-the-art systems under diverse, realistic workload conditions. \fred{We view extending and validating the framework for other agentic DAG workflows as important future work.}

\bibliographystyle{unsrt}
\bibliography{main}

\end{document}

%% file: macro.tex
\usepackage{natbib}
\usepackage{latexsym}

\usepackage{url}
\usepackage{amssymb}
\usepackage[utf8]{inputenc}
\usepackage{microtype}
\usepackage{booktabs}
\usepackage{pifont} 
\usepackage{multirow}
\usepackage{makecell}
\usepackage{paralist}
\usepackage{xspace}
\usepackage{color}
\usepackage{xcolor}
\usepackage{colortbl}
\usepackage{adjustbox}
\usepackage{hyperref} 
\usepackage[edges]{forest}
\usepackage{tikz} 
\usepackage{caption}
\usepackage{amsfonts}

\hypersetup{
    colorlinks,
    linkcolor={blue!80!black},
    citecolor={blue!80!black},
}
\tikzset{
    root/.style =             {align=center, text width=1cm, rounded corners=3pt, line width=0.3mm, fill=gray!10, draw=gray!80, font=\small},
    demographic/.style =         {align=center, text width=1.8cm, rounded corners=3pt, line width=0.3mm, fill=blue!10, draw=blue!80, font=\footnotesize},
    demographic_work/.style =    {align=center, text width=10cm, rounded corners=3pt, line width=0.3mm, fill=blue!10, draw=blue!0, font=\footnotesize},
    character/.style =         {align=center, text width=1.8cm, rounded corners=3pt, line width=0.3mm, fill=red!10, draw=red!80, font=\footnotesize},
    character_work/.style =    {align=center, text width=10cm, rounded corners=3pt, line width=0.3mm, fill=red!10, draw=red!0, font=\footnotesize},
    personalization/.style =           {align=center, text width=1.8cm, rounded corners=3pt, line width=0.3mm, fill=cyan!10, draw=cyan!80, font=\footnotesize},
    personalization_work/.style =      {align=center, text width=10cm, rounded corners=3pt, line width=0.3mm, fill=cyan!10, draw=cyan!0, font=\footnotesize},
    risk/.style =         {align=center, text width=1.8cm, rounded corners=3pt, line width=0.3mm, fill=orange!10, draw=orange!80, font=\footnotesize},
    risk_work/.style =    {align=center, text width=10cm, rounded corners=3pt, line width=0.3mm, fill=orange!10, draw=orange!0, font=\footnotesize},
}

\usepackage{CJK}

%% file: symbol_table.tex
\begin{table}[h]
    \centering
    \begin{tabular}{l | p{6cm}}
    \hline
    \textbf{Symbol} & \textbf{Description} \\
    \hline
    $Q_i$ & The $i^{th}$ end-to-end Text-to-SQL query \\
    \hline
    $q_{i,j}$ & The $j^{th}$ LLM inference requests for query $Q_i$ \\
    \hline
    $m$ & A LLM model serving instance \\

    \hline
    $\phi$ & LLM request allocation strategy \\
    \hline
    $L_{\text{in}}(q_{i,j})$ & Input length of request $q_{i,j}$ \\
    \hline
    $\hat{L}_{\text{out}}(q_{i,j})$ & Estimated output length of request $q_{i,j}$ \\
    \hline
    $t_{i,j}^m$ & Processing time of request $q_{i,j}$ on instance $m$\\
    \hline
    ${t_{\text{queue}}}_{i,j}^m$ & Expected queuing time of $q_{i,j}$ on instance $m$\\
    \hline
    ${t_{\text{comp}}}_{i',j'}^m$ & Expected compute time of $q_{i,j}$ on instance $m$\\
    \hline
    $\alpha$ & Weight balancing queuing and computing \\
    \hline
    $\alpha^*$ & Optimal value of hyperparameter $\alpha$ \\
    \hline
    $\Theta^m$ & The local queue of requests on instance $m$ \\
    \hline
    $T_i^{\text{SLO}}$ & Pre-set end-to-end SLO for query $Q_i$ \\
    \hline
    $t_{i,j}^{\text{SLO}}$ & SLO budget allocated for $q_{i,j}$ \\
    \hline
    $\tau_{\text{elapsed}}^i$ & Time elapsed since arrival of $Q_i$. \\
    \hline
    $\tau_{i,j}$ & Actual queuing delay for $q_{i,j}$ \\
    \hline
    $U_{i,j}$ & Urgency of request $q_{i,j}$ \\
    \hline
    \end{tabular}
    \caption{Summarization of notations.}
    \label{tab:symbol}
\end{table}

%% file: table1.tex

\begin{table}[htbp]
    \centering
    
    \begin{tabular}{c|c|c|c|c|c}
    \hline
    \textbf{Stage} & \textbf{Dispatcher} & \textbf{I1 (\%)} & \textbf{I2 (\%)} & \textbf{I3 (\%)} & \textbf{I4 (\%)} \\
    \hline
    \multirow{2}{*}{1} & RR & 20.8 & 23.1 & 19.9 & 26.8 \\
    \cline{2-6}
    & WB & 0.9 & 0.5 & 6.1 & 32.3 \\
    \hline
    \multirow{2}{*}{2} & RR & 10.3 & 7.8 & 10.6 & 9.6 \\
    \cline{2-6}
    & WB & 27.9 & 29.6 & 16.5 & 3.1 \\
    \hline
    \multirow{2}{*}{3} & RR & 26.5 & 23.0 & 25.5 & 21.7 \\
    \cline{2-6}
    & WB & 47.9 & 46.6 & 71.5 & 8.3 \\
    \hline
    \multirow{2}{*}{4} & RR & 42.3 & 46.0 & 44.1 & 41.8 \\
    \cline{2-6}
    & WB & 23.3 & 23.3 & 5.8 & 56.3 \\
    \hline
    \end{tabular}
    \caption{Impact of dispatching policy on task distributions. Stages 1-4 represent the aforementioned Text-to-SQL stages; I1–I4 represent different GPU instances, where I1 and I2 are A100, I3 is an A6000, and I4 is an L40 instance.}
    \label{tab:optimization-results}
\end{table}

%% file: table2.tex
\begin{table}[ht]
  \centering
  
  \begin{tabular}{l|c|c|c|c|c|c|c}
    \hline
    \textbf{Request ID} & \textbf{1} & \textbf{2} & \textbf{3} & \textbf{4} & \textbf{5} & \textbf{6} & \textbf{7} \\
    \hline
    \textbf{Arrive-at (s)} & 22.4 & 46.3 & 52.3 & 62.4 & 62.8 & 64.4 & 65.0 \\
    \hline
    \textbf{Urgency} & 14.5 & 13.2 & 19.0 & 13.1 & 19.0 & 26.9 & 21.9 \\
    \hline
  \end{tabular}
  \caption{Snapshot of a local queue before processing next LLM call. Arrive-at represents the timestamp of a LLM call arrival.}
  \label{tab:waiting-queue}
\end{table}

%% file: table3.tex
\begin{table}[h]
\centering
\caption{Overhead of alpha-tuning.}
\label{tab:overhead}
\begin{tabular}{l | r | r | r}
\hline
\textbf{Setup} & \textbf{Trace 1} & \textbf{Trace 2} & \textbf{Trace 3} \\
\hline
(Hetero-1, 0.5QPS) & 2.5s & 2.4s & 2.8s \\
\hline
(Hetero-1, 1.0QPS) & 2.7s & 2.7s & 2.9s \\
\hline
(Hetero-2, 0.5QPS) & 1.9s & 2.8s & 3.1s \\
\hline
(Hetero-2, 1.0QPS) & 2.3s & 2.7s & 3.1s \\
\hline
\end{tabular}
\vspace{-0.5em}
\end{table}